\documentstyle[11pt,aaspp4]{article}
\newcommand\beq{\begin{equation}}
\newcommand\eeq{\end{equation}}
\def\mic{{\,\mu{\rm m}}}

\begin{document}
\title{$\gamma$-ray burst afterglows as probes of galactic and
intergalactic dust}
\author{Rosalba Perna\altaffilmark{1} and Anthony Aguirre}
\affil{Harvard-Smithsonian Center for Astrophysics, 60 Garden 
Street, Cambridge, MA 02138; rperna, aaguirre@cfa.harvard.edu}
\altaffiltext{1}{Harvard Society of Fellows}
%\submitted{Accepted by The Astrophysical Journal Letters} 

\begin{abstract}

The amount and properties of high-redshift galactic and intergalactic
(IG) dust are largely unknown, but could be investigated using
multi-wavelength photometry of high-$z$ objects that have a known
intrinsic spectrum.  Observations of $\gamma$-ray burst (GRB)
afterglows appear to support the theoretical model of an adiabatic
blast wave expanding into an external medium.  In this model, the
synchrotron peak flux is independent of frequency, providing a flat
spectrum when observed over time, and therefore allowing
straightforward measurement of the relative attenuation of afterglow
flux in widely separated bands.  Applying this method to dust
extinction, we show that for a sample of afterglows which have been
corrected by galactic extinction, comparison between the number counts
of peak fluxes in $X$-ray versus optical can provide constraints on an
intergalactic component of dust.  A similar technique can probe the
redshift-dependence of extinction in GRB-forming regions without
requiring an assumed relation between extinction and reddening by the
dust.  Probing systematic changes in extinction with redshift --
particularly in IG and/or non-reddening dust -- is crucial to a proper
interpretation of the Type Ia Supernova Hubble diagram and similar
observations, and useful in understanding GRB progenitor environments.
       
\end{abstract}

\noindent {\em Subject headings:} cosmology: observations --
dust, extinction -- $\gamma$-rays: bursts

\section{Introduction}

There is now substantial evidence that $\gamma$-ray bursts (GRBs)
originate at cosmological distances (e.g. Metzger et al. 1997) from
very powerful explosions (e.g. Kulkarni et al. 1998, 1999).  They have
been detected at very high redshifts, making them useful probes of the
universe out to early cosmological epochs.  In addition, the
$\gamma$-ray emission is followed by delayed emission at longer
wavelengths, from the $X$-ray to the radio band (Costa et al. 1997;
van Paradijs et al. 1997; Frail et al. 1997).  This afterglow is
described reasonably well as synchrotron radiation, emitted when a
relativistic shell collides with an external medium (Paczy\'nski \&
Rhoads 1993; Katz 1994; Waxman 1997a,b; Wijers, Rees \& M\'esz\'aros
1997; Sari, Piran \& Narayan 1998). Afterglows, like their high energy
counterpart, are also expected to be detected out to very high
redshift. The absorption-line systems and the Ly$\alpha$ forest
visible in their spectra can therefore be used to trace the evolution
of metallicity in the universe, and to constrain, or possibly measure,
the epoch at which re-ionization of the universe occurred (e.g.  Lamb
\& Reichart 2000).

In this {\em Letter} we point out other uses that multi-wavelength
observations of afterglows might have in cosmology. Whereas it is by
now clear that GRBs (and consequently their afterglows) are far from
being standard candles, the afterglows do have a very interesting
property: the flux at the peak of the synchrotron spectrum is
independent of time for adiabatic hydrodynamic evolution (Katz 1994;
M\'esz\'aros \& Rees 1997; Sari, Piran \& Narayan 1998), and the
adiabatic shock model has received robust support from observations
(Waxman 1997a,b).  Because the frequency of the peak flux smoothly
decreases with time, this provides a strong theoretical connection
between observations at various widely-separated frequencies which is
lacking in other high-$z$ objects such as quasars or galaxies.

This property makes GRB afterglows well suited to absorption studies
of both the GRB immediate environment and host galaxy, and of the
intervening intergalactic medium (IGM).  One such possibility is the
study of intergalactic dust, which may play a role in observations of
high-$z$ Supernovae (SNe).  In the past few years, observations of
Type Ia SNe by two separate groups have revealed a progressive dimming
of SNe at high redshift with respect to the predictions of a
matter-dominated universe or even an open universe with a zero
cosmological constant (Riess et al. 1998; Perlmutter et al. 1998).
This dimming has been interpreted as evidence for acceleration in the
cosmic expansion, probably caused by a positive cosmological constant.
Two important systematic effects which must be accounted for in
drawing this conclusion are evolution in the supernova intrinsic
brightness and dust obscuration; both groups have discussed these at
length. However, their dust corrections, based on reddening, rely on
the assumption that dust has everywhere the same characteristics as in
the Milky Way, which might not be the case.  As Aguirre (1999) has
shown, a scenario in which galaxies expel a significant fraction of
their metals in winds or by radiation pressure ejection of dust, and
in which very small grains are selectively destroyed or retained by
galaxies, can self-consistently provide a viable candidate for an
intergalactic (IG) dust component which is more 'grey' (i.e. less
reddening) than dust in the Milky Way. Dust reddening properties also
vary significantly within and between galaxies.
 
GRB afterglows are observed in widely-separated bands at which even
grey dust will attenuate radiation very differently, and, as explained
above, a known relation exist between the expected fluxes in those
bands in the absence of extinction. This {\em Letter} shows that,
given a sample of GRB afterglows for which a reliable correction for
the host galaxy absorption has been made, a comparison between the
number count distribution of peak fluxes in various bands can
sensitively test the presence of a cosmological distribution of grey
dust. Whereas this test is insensitive to 
the choice of model parameters made, such as type of cosmology,
shape of the GRB luminosity function, redshift evolution of the GRB rate,
etc. (because these parameters influence in the {\em same} way
the number counts in all bands, and we are {\em
comparing} data in different bands), it does
 rely on the assumption of the constancy of the
peak flux in the various bands. In reality, several effects 
can contribute to a departure from the simple, ideal behavior
(see e.g. Meszaros, Rees \& Wijers 1998), and we discuss how
and to what extent these can be corrected for, so that our test
would still remain possible.

Besides probing the existence of IG dust, we show that the same type
of multi-wavelength study of GRB afterglows can provide a useful probe
of the evolution of metallicity (or dust-to-gas ratio) in the
GRB-forming regions of galaxies.  Independent probes of metallicity
evolution (i.e. not related to GRB sites) can then yield information
on the specific environments of GRBs in relation to galaxies, helping
to elucidate the nature GRB progenitors. Conversely, assuming a GRB
progenitor type, one could obtain information about dust in the
environment of that type of object, at very high $z$.

\section{Number counts of GRB afterglow peak fluxes}

To begin with, we adopt the simplest unbeamed synchrotron model
(Waxman 1997a,b).  Under the assumption that the magnetic field energy
density in the shell rest frame is a fraction $\xi_B$ of the
equipartition value, and that the power-law electrons carry a fraction
$\xi_e$ of the dissipated energy, the observed frequency at which the
synchrotron spectral intensity of the electrons peaks is
\begin{equation}
\nu_m(t)=2.4\times 10^{16}\left(\frac{1+z}{2}\right)^{1/2}
\left({\xi_e\over 0.2}\right)^2 
\left({\xi_B\over 0.1}\right)^{1/2} 
E_{52}^{1/2}t_{\rm hr}^{-3/2}\;{\rm Hz}\;,
\label{eq:num}
\end{equation}
where $z$ is the cosmological redshift of the source. For $\Omega=1$ and
$H_0=70~{\rm km~s^{-1}~Mpc^{-1}}$, the observed intensity at $\nu_m$ is
$
F_{\nu_m}\sim 1\;{\rm mJy}\;n_1^{1/2}\left({\xi_B\over 0.1}\right)^{1/2}
E_{52}
$
for a burst at a redshift $z\sim 1$.
Due to the combination of several effects, the afterglow flux falls off with
redshift very slowly (Lamb \& Reichart 2000; Ciardi \& Loeb 2000). For
an Einstein-De Sitter cosmology, Ciardi \& Loeb find 
$
F_{\nu_m}\propto (1+z)^{(p-3)/4}[1-(1+z)^{-1/2}]^{-2}\;,
$
where $p$ is the power-law index of the electron energy
distribution. Both the GRB and the afterglow observations are well
fitted by the value $p\approx 2.5$ (Sari, Narayan \& Piran 1998; Kumar
\& Piran 1999).  We fix the constant of proportionality so that
$F_{\nu_m}= 1$ mJy at $z= 1$.  In absence of absorption, the peak
luminosity of a given burst is then given in terms of the luminosity
distance $D_L(z)$ by $L_{\nu_m} = 4\pi D_{\rm L}^2(z)F_{\nu_m}/(1+z)$
(see e.g. Hogg 1999 and references therein).  At each redshift, we
allow for a scatter in the burst luminosity by assuming a probability
distribution that is log-normal in $L_{\nu_m}$.
%\begin{equation}
%P(L_{\nu_m},z)dL_{\nu_m} = \frac{1}{\sqrt{2\pi\sigma^2}}
%\exp\left\{-\frac{[\ln(L_{\nu_m})-\ln(L_\star)]^2}{2\sigma^2}\right\}
%{dL_{\nu_m}\over L_{\nu_m}}\;.
%\label{eq:pl}
%$\end{equation}
The mean of this distribution
%, $\langle
%L_{\nu_m}(z)\rangle=L_\star\exp\left(\sigma^2/2\right)$, 
is chosen so that
$\langle F_{\nu_m}(z)\rangle$ evolves according to Ciardi \& Loeb's relation.
We choose $\sigma=0.5$ in our computations. However, as explained in the
following, our results are not very sensitive to the particular choice of parameters.

The number (per unit time)
of GRB afterglows with peak flux above a given threshold $S$ is given by
\beq
N(>S)=\int_0^{z_{\rm max}} dz \,R(z)\,{dV_c\over dz}
\int_{L_{\rm min}(z,S)}^\infty dL_{\nu_m}
P(L_{\nu_m},z)\;.
\label{eq:nc}
\eeq
Here $R(z)$ is the GRB formation rate, which we assume to be proportional to the
star formation rate (SFR), as given by the most recent observations of Steidel 
et al. (1999), and $z_{\rm max}$ is the redshift limit of the sample.
The comoving volume is given by $dV_c=4(c/H_0)^3(1+z-\sqrt{1+z})^2(1+z)^{-7/2} 
d\Omega dz$ (e.g. Hogg 1999) in our adopted cosmology. 
The minimum luminosity that a burst at redshift $z$ must have in order
to be detected with flux $>S$ is given by 
\beq
L_{\rm min}(z,S) = \frac{4\pi D_{\rm L}^2(z) S}{(1+z)\exp\{-[\tau_{\rm dust}(z)
+\tau_{\rm abs}(z)]\}}\;.
\label{eq:lmin}
\eeq
Here $\tau_{\rm abs}$ is the optical depth to photoelectric absorption, for which
there is a contribution from the host galaxy itself and a contribution from the 
intergalactic medium (IGM). $X$-rays are not affected by photoabsorption, 
if observed at high enough frequency (see below). Optical light is
unabsorbed if observed below the Ly$\alpha$ resonance
frequency $\nu_\alpha(z)=2.47\times 10^{15}/(1+z)$~Hz, so the $V$-band 
($\nu=5.4\times 10^{14}$ Hz) and longer wavelengths will not be affected by
photoabsorption if one only considers bursts with redshifts up to $z\sim 3.5$.
We will show our results by assuming a sample with $z_{\rm max}$=3. 

The optical depth of dust at redshift $z$ is indicated by $\tau_{\rm dust}$
in Equation~(\ref{eq:lmin}). We assume
$\tau_{\rm dust}(z)=\tau_{\rm gal}(z)+\tau_{\rm grey}(z)\;,$
and consider several possibilities. 

For the grey dust component, we take the model as in Aguirre (1999):
dust is composed of equal masses of graphite and silicate spherical
grains of radii $a$, with a grain-size distribution $dN(a)/da \propto
a^{-3.5}$, $0.1\mic \le a \le 0.25\mic$.  This dust has a very flat
extinction curve blueward of the $V$ band.  We take a dust
density $\Omega_{\rm dust}(z=0) = 4.5\times 10^{-5}$, and assume that
$\Omega_{\rm dust}(z) \propto \int dt SFR$, where the $SFR$ is the
same as that assumed for the GRB rate.  Using publicly available
extinction data (see Laor \& Draine 1993) these assumptions give a
cosmologically `interesting' extinction of $\sim 0.2$ mag to $z=0.5$
in the observed $B$-band.

For the host galaxy absorption we
assume the extinction law to be the same as in our Galaxy, and model, for
an observed wavelength $\lambda$, 
\beq
A(\lambda,z) = k(z)\xi\left(\frac{\lambda}{1+z}\right)
\left(\frac{N_{\rm H}(z)}{10^{21}{\rm cm}^{-2}}
\right)\;,
\label{eq:ext}
\eeq with $\xi(\lambda)$ fitted as in Pei (1992). We then consider two
types of evolutionary scenarios for the dust-to-gas ratio $k(z)$;
model (1): k(z) = const = k(0), with k(0) = 0.78 as in the Galaxy;
model (2): k(z) traces the integrated star-formation rate. These
should represent the most extreme possibly evolutionary paths for the
dust-to-gas ratio: the second effectively assumes instantaneous mixing
of metals throughout all cosmic gas, whereas the first effectively
assumes either that cosmic metallicity does not evolve, or
alternatively that metals are so poorly mixed that while cosmic
metallicity evolves, the metallicity near the GRBs does not.  Model
(1) could better represent a scenario in which GRBs are associated
with coalescent compact objects (e.g. Eichler et al. 1989, Narayan,
Paczy\'nski \& Piran 1992), and therefore they would not be expected
to occur in particularly dense or dusty regions. On the other hand, if
GRBs are associated with the death of massive stars (e.g. Woosley
1993, Paczy\'nski 1998), then they are likely to occur in star-forming
regions.  The mean metallicity expected for such regions is
substantially higher than the cosmic mean.
This is a situation which could be more
closely described by our dust model (2)
\footnote{Note that, if GRBs are associated with star-formation
regions, then also the density in their surrounding medium is expected
to be on average higher. However, extinction is only sensitive to the
product of column density and metallicity, and therefore we 
parameterize our models with only one of them, that is the dust-to-gas ratio.}.

In Equation~(\ref{eq:ext}), $N_{\rm H}(z)$ is the average column
density intercepted by the GRB photons as they escape a host galaxy at
a redshift $z$.  To estimate this function, we have performed a Monte
Carlo calculation using the model of Ciardi \& Loeb (2000) to obtain
the probability distribution $N_H(M,z)$ for GRBs occurring (with
spatial distribution proportional to the gas density squared) in the
disks of spirals embedded in halos of mass $M$, combined with the
Press-Schechter formalism for the distribution of $dN(M,z)/dM$.  The
resulting distributions are normal in $log(N_H)$. As $z$ runs from 0
to 4, the distribution's center increases roughly linearly from 21.27
to 21.48, and its width decreases linearly from 0.915 to 0.73.\footnote{For
(GRB density) $\propto$ (gas density), the mean values are similar but
about 0.25 dex smaller.}  This should be regarded not as an accurate
and detailed model of the expected column densities (for example it
does not include the evolution of the galaxies' gas fraction), but
rather as a rough check on the possible systematic increase of GRB
host-galaxy extinction. Our two models, based on this estimate,
bracket the more detailed models of extinction evolution of Calzetti
\& Heckman (1999) and Totani \& Kobayashi (1999) which apply to the
normal stellar population in galaxy disks.

\section{The effects of dust on GRB number counts}

We present results for number count distributions with an observed
peak flux in $X$-ray, V, and J bands. Observations in the {\em hard}
$X$-ray band are necessary for a proper comparison which is not biased
by photoelectric absorption within the host galaxy. This should not be
a problem for a 5 KeV band at which {\em Swift} can observe:
Ghisellini et al. (1999) find that, at this energy, a burst at $z=0$
in a medium with solar metallicity will be unaffected by absorption
for column densities $N_{\rm H}\la 10^{23}$ cm$^{-2}$. For a burst at
$z\sim 2$ in a solar metallicity environment, the constraint becomes
$N_{\rm H}\la 10^{24}$ cm$^{-2}$. Evidence for an unusually high
column density to a burst can be found by the time-dependence of the
UV flux (Perna \& Loeb 1998), $K_\alpha$ fluorescent line emission
(Ghisellini et al. 1999), and, clearly, heavy reddening of the optical
emission. For a proper analysis such bursts should not be included in
the sample.

To probe the presence and type of dust extinction, the observed
optical-near IR bands seem the most appropriate. As explained in \S 2,
the $V$ band and longer wavelengths are unaffected by photoelectric
absorption in the IGM up to $z \sim 3.5$.  Therefore, if the 
sample is limited to redshifts lower than this, the on-board {\em
Swift} $V$ photometry could be used to determine the peak $V$ flux
(although ground-based NIR data may be necessary for obtaining the
{\em rest-frame} UBV photometry necessary for accurate de-reddening).

Figure 1 shows a comparison between number counts in $X$-ray, V, and J
bands for a sample of GRBs for which perfect corrections for Galactic
and GRB host-galaxy extinction have been made, while the grey
component (which would not be accounted for by UV/optical
de-reddening) has not been subtracted out. Figure 2 shows the same
curves, where host galaxy extinction -- quantified by models (1) and
(2) of \S 2 -- has {\em not} been corrected. In these figures, the
horizontal offset of two curves indicates the amount of extinction,
whereas the vertical {\em change} in that offset shows the evolution
of extinction with increasing redshift.

In a realistic case, one could hope to correct for host galaxy
extinction only imperfectly: there will be errors in the photometry
and $K$-correction of the rest-frame UVB fluxes, errors in determining
the appropriate extinction law ($R_V$, see Cardelli, Clayton \& Mathis
1989), and variations in the extinction law even for a given
$R_V$. The extinction may also be time-dependent (Waxman \& Draine,
1999).  However, we find that even if the correction leaves a
`residual', the statistical difference is still readily
measurable. We performed a Monte Carlo simulation of data
drawn from the $X$-ray distribution and of data drawn from the
$V$-band distribution corrected for galactic extinction but not for a
grey intergalactic component (as in Figure 1).  To each flux was assigned a
fractional error that was randomly generated from a Gaussian
distribution of width $\sigma_{\rm err}$. We took $\sigma_{\rm
err}=0.5$, which corresponds to about half to 1 magnitude error in
extinction correction, larger than the quoted errors on $A_v$ for GRBs
observed so far (see e.g. Bloom et al. 1998, Vreeswik et al. 1999).
By performing a Kolmogorov-Smirnov test (see e.g. Press et al. 1992)
with the two sets of simulated data, we found that with $N\ga 70$
bursts the two distributions can be distinguished with a confidence
level $\sigma\ga 3$. 

Note that, if the host galaxy extinction is decreasing with $z$ (as in
our model (1), or as in Calzetti \& Heckman 1999 and Totani \&
Kobayashi 1999), then galactic extinction would never mimic the
effects of IG grey dust. On the other hand, if the likely scenario is
as in model (2), then our ability to distinguish IG dust from
increasing host galaxy extinction would depend on the accuracy with
which host galaxy dust correction can be made.

A very clear distinction can also be made between the distributions
described by model (1) and (2) in Figure 2. A random simulation as
described above showed that with $N\ga 20$ bursts the two $V$-band
distributions in the case where there is no grey dust component can be
distinguished with a confidence level $\sigma\ga 3$. Similar results
hold for the corresponding distributions which also include grey dust.
In these simulations we have assumed the same $\sigma_{\rm err}$ as
above. However, when no correction for galactic extinction is
required, this error is likely to be much smaller.  This study is
therefore very sensitive to the conditions (i.e. metallicity and
density) in the environment of GRBs. Combining such analysis with
independent probes of the metallicity and density evolution of the
medium in galaxies and in the IGM\footnote{Spectra of the GRBs
themselves could also be used to such purposes, as shown by Lamb \&
Reichart (2000).}, significant information can therefore be obtained
about the location of GRBs within galaxies.  This is particularly
relevant in order to distinguish scenarios in which GRBs are due to
coalescence of two compact objects with respect to scenarios in which
GRBs are associated with the death of massive stars.  Since reddening
and extinction are measured rather independently, the technique can
also give information about the reddening properties of dust in GRB
progenitor environments.

We need to emphasize once more that our method is rather insensitive
to the choice of model parameters made, such as type of cosmology,
shape of the GRB luminosity function, GRB evolution with redshift,
etc.  This is because these parameters influence in the {\em same} way
the number counts in all bands, and our method is based on a {\em
comparison} of data in different bands.  What can change (due for
example to a different distribution of the bursts with the redshift)
are statistical quantities such as the number of bursts required to
distinguish between two models with a given confidence level, but we
found the difference in this number required for various models to be
rather marginal.  The main uncertainty of this method lies in the
fact, already discussed in \S 1, that the perfect, adiabatic model
with a constant peak flux is a clear approximation that does not take
into account several effects which might cause a departure from this
simple behavior. Meszaros et al.  found that a number of factors
(such as angular anisotropy of the fireball, properties of the
environment, later re-energization of the afterglow, etc.)  can lead to
either an increase or to a decrease of the peak flux with time,
depending on what the particular conditions (such as for example
density gradients in the external medium) are for a given burst. This
means that, most likely, such disturbances should not give rise to
systematic effects, but rather would lead to variations equivalent to
having a larger scatter in the errors on the fluxes.  
If this is indeed the case, the effects produced by grey dust should
not be easily washed out: a random simulation of data with errors in
peak fluxes drawn from a distribution with $\sigma_{\rm err}$=2
(i.e. a fractional error of 200\%) showed that the distributions in
Figure 1 can still be distinguished with a confidence level
$\sigma\ga3$ if the number of afterglows is $\ga 200$. Clearly, the
situation would become more difficult if these disturbing effects,
instead of being stochastic, had a trend which would systematically
reduce the emission in one (or more) bands with respect to the
others. However, unless these hypothetical systematics have also an
evolution with redshift, it should be possible to correct for them by
measuring ratios between peak fluxes in various bands for a sample of
bursts at low $z$ (which would only be marginally affected by IG
dust). A systematic decrease in the flux at peak frequency could also
mimic the difference between a dusty and a relatively dust-free 
GRB environment, but again only at one redshift; {\em evolution} in the
GRB environment should still be measurable. 
Such a study, to be statistically significant, would need a
large number of observed afterglows. This should be possible with {\em
Swift}, which will observe $\sim 300$ bursts per year, and will
monitor their emission from the $X$-ray to the V band (and other
instruments will very likely obtain NIR photometry of the
afterglows). It will be an ideal tool for such a study.

\section{Conclusions}

One of the outstanding problems in observational cosmology is to
understand the type and amount of dust existing at high-$z$, whether
outside galaxies or within them. Traditional methods of estimating
extinction require a known dust reddening curve and an accurate model
for the intrinsic UV/optical spectrum of the observed object, both of which may
be rather uncertain; very few existing techniques allow the measurement of
absolute extinction.

In this {\em Letter}, we have proposed GRB afterglows as probes of
dust in cosmology. Like QSOs and galaxies they can be observed to very
high $z$.  In addition, there is a well-based theoretical spectrum
which is flat between $X$-ray and near infra-red frequencies. This
allows very straightforward estimates of dust extinction {\em
independent of its reddening properties} (though the reddening would
be observable and useful) in GRB environments or in the intergalactic
medium; the former can help help determine the properties of GRB
progenitor environments (including the dust properties), while the
latter is crucial for careful interpretation of the Type Ia
Supernova Hubble diagram.

While relying on the assumption that afterglow evolution is well
described by the simplest adiabatic blast model, with a large number
of afterglows (such as {\em Swift} will provide), the technique should
be applicable even if deviations from the theory occur.

\newpage

\begin{figure}[t]
\centerline{\epsfysize=5.7in\epsffile{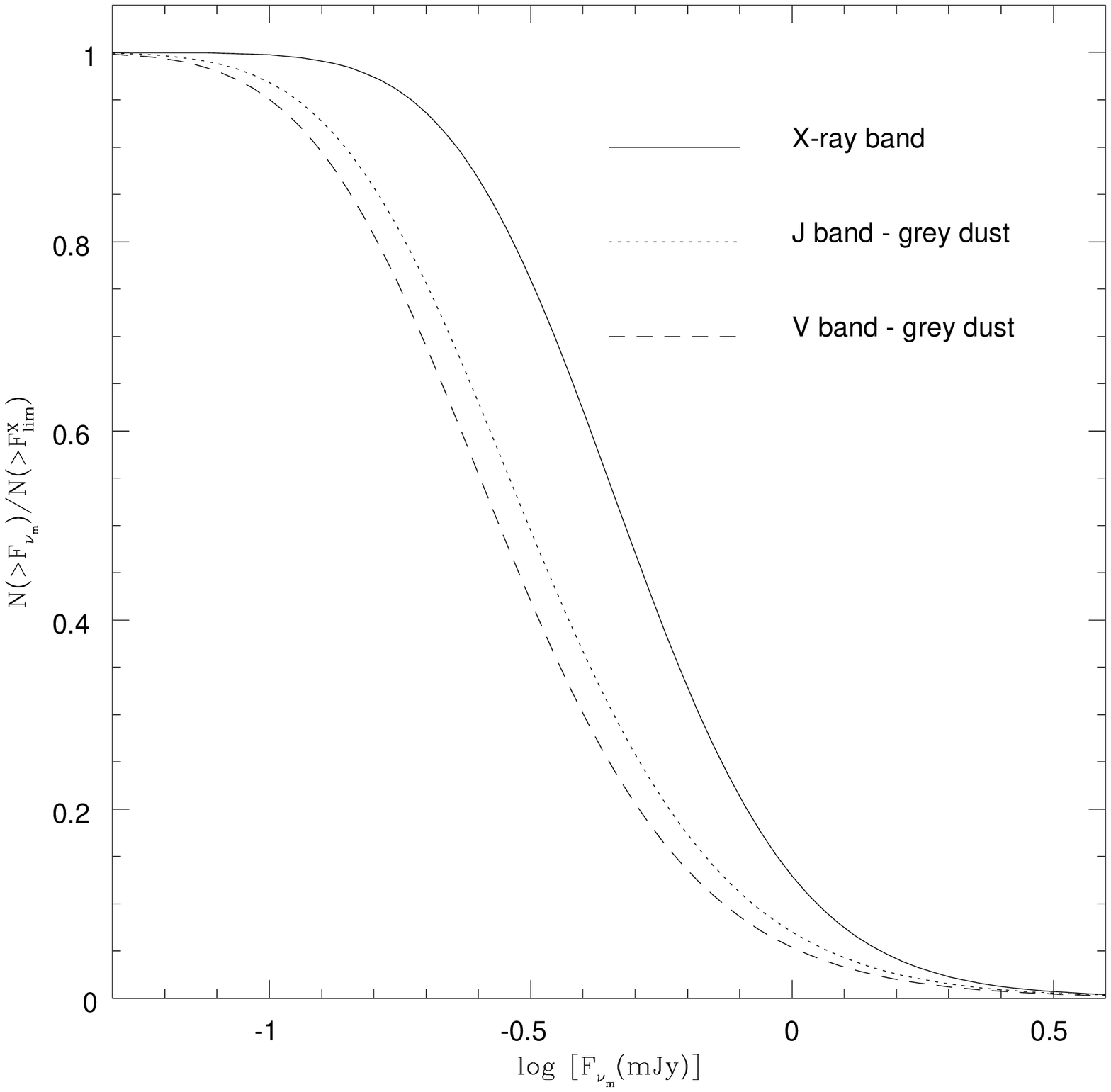}}
\caption{Cumulative number count distribution of GRB afterglow peak fluxes
in the $X$, J, and V bands 
for a sample corrected by galactic extinction but not by the extinction 
due to an intergalactic grey dust component.}
\label{fig:1}
\end{figure}

\begin{figure}[t]
\centerline{\epsfysize=5.7in\epsffile{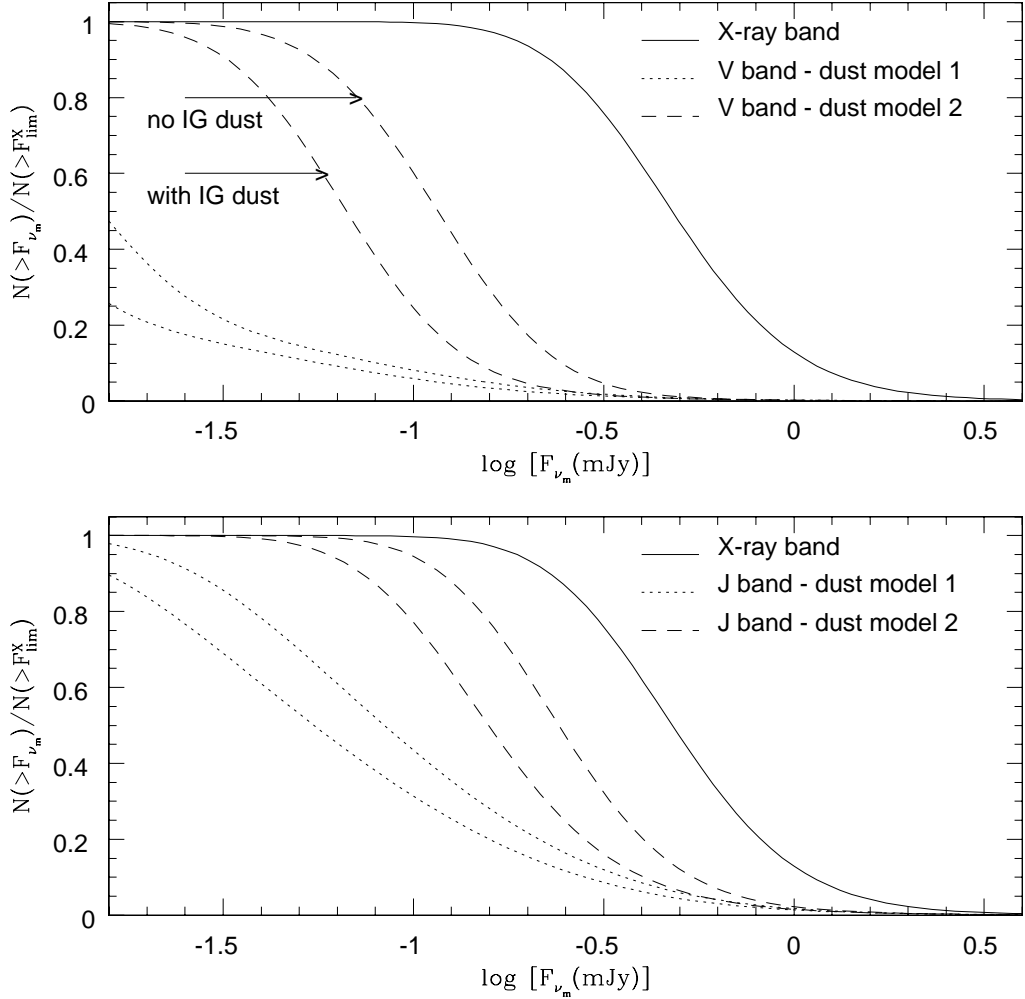}}
\caption{Cumulative number count distribution of GRB afterglow peak fluxes 
in the $X$, and V bands
(upper panel) and the $X$ and J band (lower panel). In both panels, the dust models (1)
and (2) refer to the host galaxy dust models
 described in the text. For each model, the upper curve
has only the component from galactic extinction, while the lower curve includes 
also an intergalactic grey dust component.}
\label{fig:2}
\end{figure}

\end{document}